\documentclass[sigconf]{acmart}
\AtBeginDocument{%
  }

\copyrightyear{2026}
\acmYear{2026}
\setcopyright{cc}
\setcctype{by}
\acmConference[KDD '26]{Proceedings of the 32nd ACM SIGKDD Conference on Knowledge Discovery and Data Mining V.2}{August 09--13, 2026}{Jeju Island, Republic of Korea}
\acmBooktitle{Proceedings of the 32nd ACM SIGKDD Conference on Knowledge Discovery and Data Mining V.2 (KDD '26), August 09--13, 2026, Jeju Island, Republic of Korea}
\acmDOI{10.1145/3770855.3818986}
\acmISBN{979-8-4007-2259-2/2026/08}

\usepackage[dvipsnames]{xcolor}

\newcommand{\mmt}{\textsc{TokaMind}}
\newcommand{\bench}{{\textsc{TokaMark}}}
\newcommand{\DatasetSize}{11,573}
\newcommand{\TotalSignals}{39}

\usepackage{acro}

\DeclareAcronym{fm}{
  short=FM,
  long=Foundation Model
}

\usepackage{soul}
\usepackage{pifont}

\newcommand{\cmark}{\textcolor{green!60!black}{\ding{51}}} 
\newcommand{\xmark}{\textcolor{red!70!black}{\ding{55}}}   

\usepackage{array}
\usepackage{enumitem}

\usepackage{booktabs}
\usepackage{longtable}
\usepackage{multirow}
\usepackage{adjustbox}
\usepackage{bm}
\usepackage{stfloats}

\setcopyright{none}
\renewcommand\footnotetextcopyrightpermission[1]{}

\setcopyright{none}
\renewcommand{\shortauthors}{}

\setcopyright{none}

\makeatletter
\gdef\@received{}
\gdef\@revised{}
\gdef\@accepted{}
\makeatother

\begin{document}

\makeatletter
\let\@oddhead\@empty
\let\@evenhead\@empty
\let\@oddfoot\@empty
\let\@evenfoot\@empty
\makeatother



\title{TokaMind: A Multi-Modal Transformer Foundation Model for Tokamak Plasma Dynamics}






\author{Tobia Boschi}
\affiliation{%
  \institution{IBM Research}
  \city{Dublin}
  \country{Ireland}}
\email{tobia.boschi@ibm.com}

\author{Andrea Loreti}
\affiliation{%
  \institution{UK Atomic Energy Authority}
  \city{Abingdon}
  \country{UK}}
\email{andrea.loreti@ukaea.uk}

\author{Nicola C. Amorisco}
\affiliation{%
  \institution{UK Atomic Energy Authority}
  \city{Abingdon}
  \country{UK}}
\email{nicola.amorisco@ukaea.uk}

\author{Rodrigo H. Ordonez-Hurtado}
\affiliation{%
  \institution{IBM Research}
  \city{Dublin}
  \country{Ireland}}
\email{rodrigo.ordonez.hurtado@ibm.com}

\author{Cécile Rousseau}
\affiliation{%
  \institution{IBM Research}
  \city{Dublin}
  \country{Ireland}}
\email{rousseau.cecile@ibm.com}

\author{George K. Holt}
\affiliation{%
  \institution{UK Atomic Energy Authority}
  \city{Abingdon}
  \country{UK}}
\email{george.holt@ukaea.uk}

\author{Eszter Sz\'ekely}
\affiliation{%
  \institution{UK Atomic Energy Authority}
  \city{Abingdon}
  \country{UK}}
\email{eszter.szekely@ukaea.uk}

\author{Alexander Whittle}
\affiliation{%
  \institution{UK Atomic Energy Authority}
  \city{Abingdon}
  \country{UK}}
\email{alexander.whittle@ukaea.uk}

\author{Samuel Jackson}
\affiliation{%
  \institution{UK Atomic Energy Authority}
  \city{Abingdon}
  \country{UK}}
\email{samuel.jackson@ukaea.uk}

\author{Adriano Agnello}
\affiliation{%
  \institution{STFC Hartree Centre}
  \city{Daresbury}
  \country{UK}}
\email{adriano.agnello@stfc.ac.uk}

\author{Stanislas Pamela}
\affiliation{%
  \institution{UK Atomic Energy Authority}
  \city{Abingdon}
  \country{UK}}
\email{Stanislas.Pamela@ukaea.uk}

\author{Alessandra Pascale}
\affiliation{%
  \institution{IBM Research}
  \city{Dublin}
  \country{Ireland}}
\email{apascale@ie.ibm.com}

\author{Robert Akers}
\affiliation{%
  \institution{UK Atomic Energy Authority}
  \city{Abingdon}
  \country{UK}}
\email{rob.akers@ukaea.uk}

\author{Juan Bernabe-Moreno}
\affiliation{%
  \institution{IBM Research}
  \city{Dublin}
  \country{Ireland}}
\email{juan.bernabe-moreno@ibm.com}

\author{Vassil Alexandrov}
\affiliation{%
  \institution{STFC Hartree Centre}
  \city{Daresbury}
  \country{UK}}
\email{vassil.alexandrov@stfc.ac.uk}

\author{Mykhaylo Zayats}
\affiliation{%
  \institution{IBM Research}
  \city{Dublin}
  \country{Ireland}}
\email{mykhaylo.zayats1@ibm.com}

\renewcommand{\shortauthors}{Boschi et al.}


\begin{abstract}

We present \textbf{\mmt}, to our knowledge the first open-source foundation model for tokamak plasma dynamics, based on a Multi-Modal Transformer (MMT) and pretrained on heterogeneous diagnostics from the publicly available MAST dataset.
\mmt{} supports multiple data modalities (time-series, 2D profiles, and videos) with different sampling rates, robust missing-signal handling, and efficient task adaptation via selectively loading and freezing four model components.
To represent multi-modal signals, we use a lightweight fixed-basis Discrete Cosine Transform embedding (DCT3D) and provide a clean interface for alternative embeddings (e.g., Variational Autoencoders).
We evaluate \mmt{} on the recently introduced MAST benchmark \bench{}, which comprises 14 tasks with heterogeneous reconstruction and forecasting objectives.
Our results show that fine-tuned \mmt{} outperforms the strongest benchmark baseline on all but one task. Compared with training the same architecture from scratch under a matched epoch budget, warm-start adaptation is most beneficial on demanding downstream settings, including long-horizon forecasting and high-dimensional equilibrium objectives.
These findings highlight the value of multi-modal pretraining for tokamak plasma dynamics and provide a practical, extensible foundation for future fusion modeling tasks.
Training code and model weights are publicly available at \href{https://github.com/UKAEA-IBM-STFC-Fusion-FMs/tokamind}{github.com/UKAEA-IBM-STFC-Fusion-FMs/tokamind} and \href{https://huggingface.co/UKAEA-IBM-STFC}{huggingface.co/UKAEA-IBM-STFC}, respectively.

\end{abstract}








\maketitle

\section{Introduction}
\label{sec:intro}

\paragraph{Fusion energy and tokamak operation}
Magnetic-confinement fusion aims to produce abundant, low-carbon energy by sustaining high-temperature plasmas under tight stability and safety constraints \cite{walker2020introduction}.
Although tokamaks are among the leading concepts for magnetic-confinement fusion, achieving reliable, high-performance operation requires accurate reconstruction and forecasting of plasma behavior, which underpins monitoring and, ultimately, control across operating regimes \cite{degrave2022magnetic}.
Recent national guidelines such as the Fusion Science \& Technology Roadmap~\cite{usdepofenergy2025roadmap} emphasize accelerating progress toward a demonstration of a practical fusion power plant by advancing integrated modeling and AI-enabled, data-driven methods for plasma analysis and prediction.

\paragraph{Tokamak data challenges}
Tokamak experiments are governed by strongly-coupled, nonlinear plasma dynamics \cite{anirudh20232022} and produce heterogeneous signals spanning multiple modalities and time scales (e.g., scalar time-series, structured profiles, and imaging).
Because the plasma state is not directly observable, it must be inferred from indirect and noisy measurements \cite{costley2001iter}.
This makes reconstruction and forecasting intrinsically challenging and often ill-posed, particularly when diagnostic availability varies across shots or operating regimes \cite{joung2023gs}.
Beyond reconstruction of the instantaneous plasma state, modeling plasma evolution also requires actuator information (e.g., fueling commands and voltages applied by power supplies), which provides essential context for how the system is driven \cite{felici2012non}.
Finally, experimental datasets commonly include missing channels and dropouts and induce task-dependent input/output sets, posing a practical challenge for methods trained on real experimental records \cite{jalalvand2025multimodal}.

\paragraph{Related work: current approaches and limitations}
The above challenges have motivated growing interest in data-driven pipelines that complement physics-based tools and accelerate analysis and control in practical experimental settings \cite{humphreys2020advancing}.
Recent Machine Learning approaches have demonstrated promising performance on targeted reconstruction and control-oriented tasks \cite{shousha2023machine, kim2026real}, typically by learning mappings from a selected set of diagnostic inputs (and sometimes actuators) to task-specific outputs.
However, many existing models are specialized to a particular objective, time horizon, and curated signal set, and assume fixed input/output schemas and consistent diagnostic availability \cite{zheng2023disruption}.
This specialization limits reuse across tasks with different targets, reduces robustness to missing channels and dropouts, and hinders transfer to new devices or operating regimes.
In short, this motivates more generalist approaches that (i) learn transferable representations of plasma dynamics directly from heterogeneous data, (ii) support a wide range of downstream objectives with minimal task-specific adaptation, and (iii) generalize across devices and operating regimes, particularly in low-data settings.

\paragraph{Foundation models for plasma dynamics}
A natural way to address these constraints is to move from task-specific models to \emph{foundation models} (FMs): models \emph{pretrained} on broad, heterogeneous data and objectives so that a single initialization can be efficiently adapted to many downstream tasks and changing signal schemas \cite{awais2025foundation, fei2022towards}.
In this paradigm, pretraining aims to learn transferable representations that reduce the amount of task-specific data and tuning required, and can improve robustness under changing inputs, targets, and operating conditions.
In fusion, early perspectives have begun to articulate how FM-style pretraining could support experimental workflows \cite{churchill2025ai}.
The adoption of domain-specific FMs is increasingly practical also thanks to open data infrastructure and standardized benchmarks: FAIR-MAST provides access to data from the MAST experiment \cite{sykes2001first, counsell2005overview, meyer2009overview}, while \bench{} standardizes multi-task evaluation \cite{jackson2024fair, jackson_open_2025, rousseau2026tokamark}.

\paragraph{TokaMind: a multi-modal transformer FM for fusion plasma}
Motivated by the need for transferable, schema-flexible models of plasma dynamics, we introduce \textbf{\mmt{}}, to our knowledge the first open-source foundation model for tokamak plasma data based on a \textbf{Multi-Modal Transformer (MMT)} \emph{pretrained} on the MAST dataset \cite{jackson2024fair, jackson_open_2025}.
Through broad multi-signal pretraining, \mmt{} yields a reusable initialization that can be efficiently adapted to new input/output schemas and heterogeneous reconstruction and forecasting tasks.
Efficient codec-based signal compression, coupled with a modular design, yields a lightweight model (\mbox{$<7$M} parameters) that supports heterogeneous modalities and time scales under changing signal availability.

Our main contributions are:
\begin{itemize}[leftmargin=2em]
    \item \textbf{A schema-flexible, multi-modal transformer framework} for tokamak data (time-series, profiles, and videos) with robust missing-signal handling.
    \item \textbf{A modular tokenization and codec interface} that converts windowed multi-rate signals into a variable-length set of tokens via chunking and modality-aware embeddings, with a strong default fixed-basis DCT3D codec and clean hooks for learned alternatives (e.g., Variational Autoencoders).
    \item \textbf{Efficient adaptation mechanisms} (warm-start + selective freezing) that reuse pretrained components across tasks with diverse objectives.
    \item \textbf{Benchmark validation}, showing that fine-tuned \mmt{} improves over the strongest \bench{} baseline on nearly all tasks. Comparisons against matched training from scratch show that pretraining is most valuable for demanding downstream settings---supporting the value of transferable representations of plasma dynamics.
\end{itemize}

\paragraph{Evaluation settings.}
We validate \mmt{} on \bench{}~\cite{rousseau2026tokamark}, a standardized MAST benchmark comprising 14 tasks with curated objectives, preprocessing, and evaluation protocols, together with statistical and neural baselines.

\paragraph{Paper outline.}
The remainder of the paper is organized as follows.
Section~\ref{sec:data_and_tasks} summarizes the benchmark setting;
Section~\ref{sec:tokenization} describes \mmt{} tokenization and embedding codecs;
Section~\ref{sec:model} details the \mmt{} MMT architecture;
Section~\ref{sec:training_adaptation} presents training and adaptation;
Section~\ref{sec:experiments} reports experimental results and ablation studies;
Section~\ref{sec:limitations} discusses limitations and ethical considerations; and
Section~\ref{sec:conclusion} provides concluding remarks and outlines future research directions.

\section{Benchmark Overview: Data and Tasks}
\label{sec:data_and_tasks}

For benchmarking purposes, we use \bench{}, a recently proposed benchmark designed to standardize the development and evaluation of AI models for fusion plasma dynamics \cite{rousseau2026tokamark}.
Importantly, \bench{} moves beyond individual tasks and instead presents a suite of interconnected scientific objectives curated to probe core capabilities required of AI models for fusion plasmas: (i) representation learning from complex and incomplete data; (ii) reasoning across diverse timescales; (iii) robustness to missing information; and (iv) generalization across operating regimes. From a technical point of view, \bench\ provides harmonized access to multi‑modal diagnostics from MAST tokamak experiments and formalizes preprocessing steps such as temporal windowing, signal alignment, and metadata normalization. The use of \bench\ enables structured and reproducible training pipelines and, crucially, transparent and consistent comparison between different models. It also provides a suite of baseline models and unified evaluation tooling, ensuring that new approaches can be benchmarked fairly and rigorously. For our purposes, we adhere to the taxonomy and protocols of \bench\ as closely as possible.

\subsection{Data Summary}
\label{subsec:data_summary}

Based on the FAIR-MAST system and its associated MAST dataset~\cite{jackson_open_2025, jackson2024fair}, \bench{} provides task-configured data loaders with streamlined access to the signals required for each task.
The benchmark includes data for \DatasetSize{} \emph{shots} (or \emph{discharges}) of short duration---2--3~s experiments, typical of tokamak operation---and makes available \TotalSignals{} signals spanning a range of diagnostic categories.
These include magnetic measurements (flux loops, pickup coils, Mirnov coils), radiative diagnostics (D$_\alpha$, soft X‑ray channels), kinetic measurements (Thomson scattering and interferometry), actuator commands (voltages applied to solenoid and poloidal field coils, gas fueling, neutral beam injection), and derived equilibrium quantities such as shape parameters, plasma boundaries, $J_{\text{tor}}$ metrics, and flux maps.

In addition to physical categories, the signals are distinguished by their sampling frequency (ranging from 0.2~kHz up to 500~kHz) and structural shape, including:
\begin{itemize}[leftmargin=2em]
    \item \textbf{time series}, represented as 1D tensors (scalar value over time);
    \item \textbf{profiles}, represented as 2D tensors (vector over time);
    \item \textbf{videos}, represented as 3D tensors (2D images over time).
\end{itemize}
Together, these heterogeneous signals constitute a rich, multi-rate, and multi-modal experimental dataset which, as any other, also contains gaps and noise.

\subsection{Tasks Definition}
\label{subsec:task_definition}

\bench\ defines a suite of 14 supervised \emph{window-to-window} learning tasks organized into 4 groups, each representing a broad scientific objective: instantaneous equilibrium reconstruction, fast magnetics dynamics, profile dynamics, and prediction of MHD activity.
%
Each \bench~ task specifies three (possibly overlapping) sets of signals:
\begin{enumerate}[label=(\roman*)]
    \item an \emph{input diagnostic set} $\mathcal I$ (sensor measurements),
    \item an optional \emph{actuator set} $\mathcal A$ (machine control parameters),
    \item an \emph{output set} $\mathcal O$ (targets to be reconstructed or predicted),
\end{enumerate}
together with a windowed sample defined by four times $t_0 \le t_1 \le t_2 \le t_3$.
Inputs are observed over the context window $[t_0,t_1]$, while outputs are defined over the target window $[t_2,t_3]$.
When actuators are included, they are provided over $[t_0,t_3]$, covering both context and target horizon.


Depending on the objective, tasks fall into one of three families:
\begin{enumerate}[label=(\roman*)]
    \item \emph{reconstruction:} $t_2=t_0$ and $t_3=t_1$, i.e., targets are defined on the same window as the inputs;
    \item \emph{autoregressive forecasting:} $t_2\ge t_1$ and $t_3>t_2$ with $\mathcal{O}\subseteq \mathcal{I}$, i.e., future values of (a subset of) input signals are predicted; 
    \item \emph{reconstructive forecasting:} $t_2\ge t_1$ and $t_3>t_2$ with outputs in $\mathcal{O}$ that may differ from $\mathcal{I}$, i.e., related diagnostics or derived quantities are predicted over a future horizon.
\end{enumerate}
Formally, each task can be described by the following window-to-window prediction problem:
\begin{equation*}
\label{eq:problem_definition}
\hat{y}_{O}([t_2,t_3]) \;=\; f_{\theta}\!\left(x_{I}([t_0,t_1]),\, u_{A}([t_0,t_3])\right).
\end{equation*}
In addition, \bench\ distinguishes \emph{Markovian} and \emph{non-Markovian} settings, depending on whether the target can be predicted from a short input window or if a longer temporal history is required.

For the formal descriptions of tasks, group-level organization, and full benchmark details, we refer the reader to \cite{rousseau2026tokamark}.

\section{\mmt{} Tokenizer}
\label{sec:tokenization}

\begin{figure*}
    \centering
    \includegraphics[width=\textwidth]{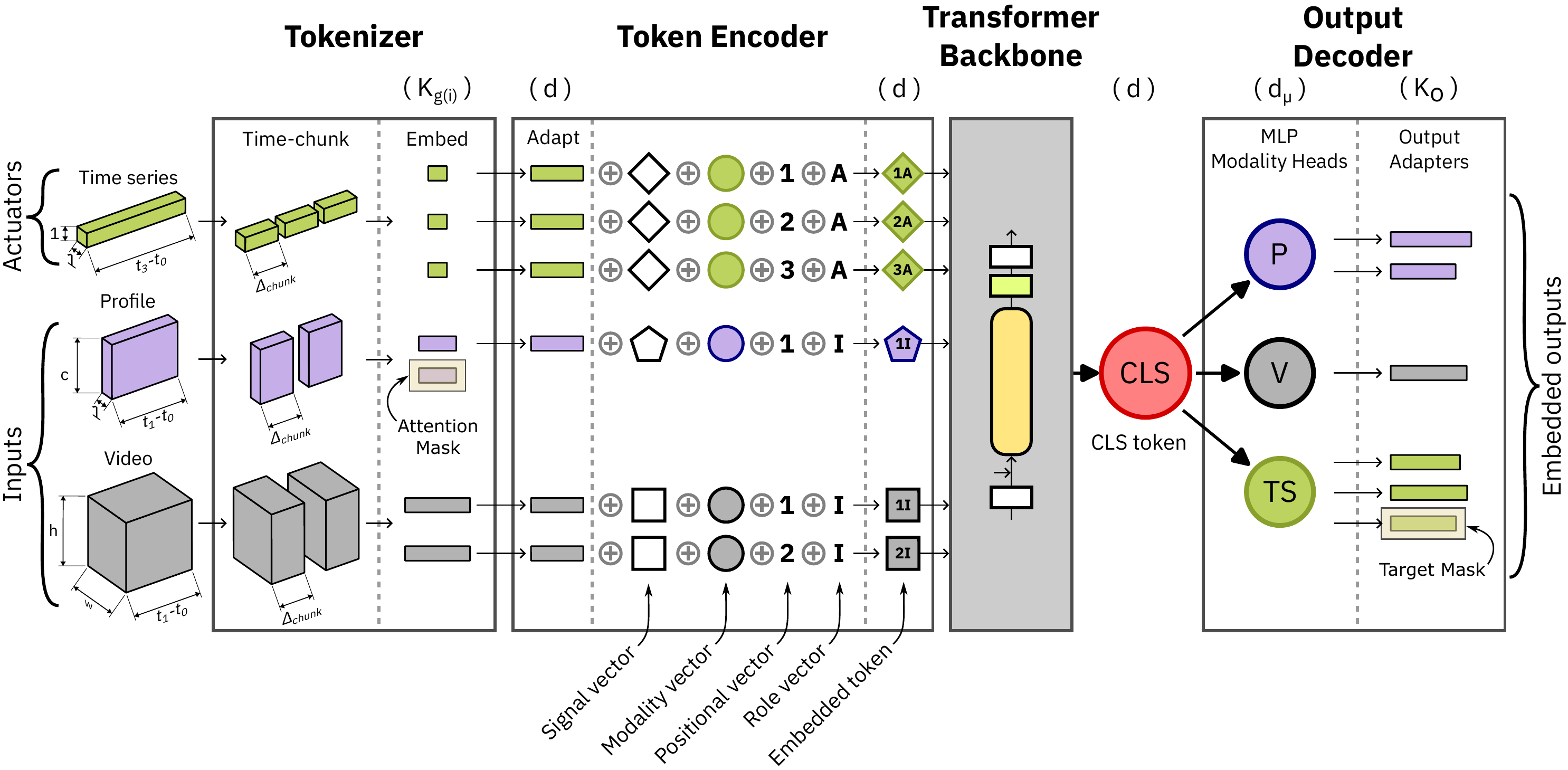}
    \captionsetup{skip=2pt}
    \caption{TokaMind tokenization and model architecture. Windowed multi-modal inputs $\mathcal I$ and actuators $\mathcal A$ are chunked and embedded by signal-specific codecs $E_g$ to produce token embeddings $z_i\in\mathbb{R}^{K_{g(i)}}$ (outputs are embedded at window level to form targets in the same space). A Token Encoder projects each token to the shared model dimension $d$ and adds learned metadata embeddings (signal, role, modality, relative position). A Transformer Backbone processes the variable-length token set using an attention mask for missing/padded tokens and outputs the \texttt{[CLS]} (classification) token embedding. Modality-specific heads (TS: time series, P: Profile, V: Video) and per-output adapters predict embedded targets $\hat{y}_o\in\mathbb{R}^{K_o}$; a target-availability mask $m_o$ excludes missing outputs from the supervised loss.}
    \label{fig:architecture}
\end{figure*}

The first step performed by \mmt{} (see Fig.~\ref{fig:architecture}) is the tokenization process that converts windowed multi-modal signals with heterogeneous sampling rates into the model tokens.
Given a windowed sample defined by $(\mathcal{I},\mathcal{A},\mathcal{O})$ and both time windows $[t_0,t_1]$ and $[t_2,t_3]$, tokenization produces a variable-length set of token embeddings $\{z_i\}_{i=1}^{L}$, where each token represents a chunk of signal data with $L$ being the number of chunks used to decompose each window.

\subsection{Transform Chain}
\label{sec:window_processing}

For each window, tokenization proceeds through the following steps:
\begin{enumerate}[leftmargin=2em]
    \item \textbf{Chunking:} Each input and actuator signal window is decomposed into fixed-duration chunks of length $\Delta_{\text{chunk}}$ with stride $s$ (where $s=\Delta_{\text{chunk}}$ for non-overlapping chunks).
    Chunking is a role-specific action: input chunks cover the window $[t_0,t_1]$, while actuator chunks may span a longer interval $[t_0,t_3]$---for forecasting tasks, they cover both the context and target horizon. Output signals are \emph{not} chunked.
    Also, chunking is based on fixed-size time intervals, so signals with different sampling rates contribute different numbers of samples per chunk while sharing the same chunk grid (and thus the same number of chunks prior to filtering).
    Chunks constitute the model's basic input elements and enable flexible context: the amount of history provided to the model is controlled by retaining more or fewer recent chunks, without changing the architecture.

    \item \textbf{Validity selection:} Partially observed signals are retained, while empty or fully missing signals are marked as unavailable. Windows are discarded if they do not contain at least one valid input/actuator and at least one valid output target.
    Optionally, windows are subsampled per shot by enforcing a minimum spacing in $t_1$.
    In order to bound the number of elements fed to the model, for each role we retain at most the $M$ most recent valid chunks (i.e., those closest in time to the end of the prediction interval at $t_3$).

   \item \textbf{Embedding generation and missing-value handling:} Each retained input/actuator chunk is transformed into a token embedding $z_i \in \mathbb{R}^{K_{g(i)}}$ using a signal-specific embedding codec, where $g(i)$ denotes the signal associated with the $i$-th token.
   Outputs are transformed at window level to provide supervision targets in the same embedding space.
   Before codec tuning or embedding generation, remaining non-finite values are handled using one of three strategies: (i) zero filling; (ii) temporal interpolation followed by spatial interpolation and a zero fallback; or (iii) no imputation when the codec can natively accept non-finite values.
\end{enumerate}

\noindent
After tokenization, each window yields:
(i) a variable-length set of token embeddings $\{z_i\}_{i=1}^{L}$ for the available input/actuator chunks, and
(ii) window-level embedded targets for each output signal used for supervision, together with indicators of which outputs are present.

In the remainder of this section, we describe the embedding codecs $E_g$ used to map native-rate chunks (and window-level outputs) into the shared embedding space.

\subsection{Chunk and Window Embedding}
\label{subsec:embeddings}

The embedding layer maps heterogeneous modalities into a common representation space, enabling a uniform token interface and a consistent supervision mechanism across tasks.
For each signal $g$, we define an embedding codec $E_g$ that maps an input chunk window (or an output window) $x$ to a vector in $\mathbb{R}^{K_g}$.
Applied to input/actuator chunks, this yields a set of token embeddings $z_i = E_{g(i)}(x_i) \in \mathbb{R}^{K_{g(i)}}$, while output signals are embedded at the \emph{window level} to form supervision targets $y_o = E_o(x) \in \mathbb{R}^{K_o}$ in the same representation space.

Our default embedding choice is the Discrete Cosine Transform codec (DCT3D) \cite{boussakta2004fast}, used for pretraining and all downstream tasks.
However, it is worth noting that the \mmt{} framework is flexible and embedding-agnostic, and we additionally support identity embeddings (no compression) and learned embeddings produced by Variational Autoencoders (VAEs). 
As part of our experimental results in Section~\ref{sec:experiments}, we evaluate VAE embeddings in an ablation study on a subset of \bench{} tasks.

\subsubsection{\textbf{DCT3D}}
\label{subsubsec:dct3d}
DCT3D, short for the three-dimensional discrete cosine transform \cite{boussakta2004fast}, is a lightweight codec that compresses time-dependent signals by projecting them onto an orthonormal cosine basis over \emph{space/channel} and \emph{time}.
A chunk (or output window) is represented as an array $x$ that can be a 1D time series, a 2D profile, or a 3D map/video.
We treat all cases uniformly by reshaping $x$ to a 3D tensor $x^{3D}\in\mathbb{R}^{H\times W\times T}$, using singleton dimensions when needed (e.g., $(T)\!\to(1,1,T)$ and $(C,T)\!\to(C,1,T)$).
Let $D_H\in\mathbb{R}^{H\times H}$, $D_W\in\mathbb{R}^{W\times W}$, and $D_T\in\mathbb{R}^{T\times T}$ denote the orthonormal DCT-II bases along each axis. Applying the separable 3D DCT-II yields the coefficient tensor
\[
\mathcal{C} = \mathrm{DCT}(x^{3D}) \in \mathbb{R}^{H\times W\times T},
\]
and, equivalently,
\[
\mathrm{vec}(\mathcal{C}) = \left(D_T \otimes D_W \otimes D_H\right)\,\mathrm{vec}\big(x^{3D}\big),
\]
where $\mathrm{vec}(\cdot)$ stacks tensor entries into a vector and $\otimes$ denotes the Kronecker product.

To form a compact representation for signal $g$, we retain a signal-specific ordered set of coefficient indices
\[
\mathcal{I}_g \subseteq
\{1,\ldots,H\}\times\{1,\ldots,W\}\times\{1,\ldots,T\}.
\]
The embedding is then the vector of selected coefficients:
\[
z = \left[\mathcal{C}_{hwt}\right]_{(h,w,t)\in\mathcal{I}_g}
\in \mathbb{R}^{K_g},
\qquad K_g = |\mathcal{I}_g|.
\]
Unlike a fixed low-frequency truncation block, the retained index set can include higher-frequency coefficients whenever they carry relevant signal energy.
Reconstruction scatters the selected coefficients into an otherwise zero-filled tensor with shape $(H,W,T)$ and then applies the inverse transform (IDCT) to recover $\hat{x}^{3D}$.

\paragraph{Main benefits.}
DCT3D provides three practical benefits.
First, it yields a fixed-size embedding controlled by $K_g$, enabling a uniform token representation across modalities and sampling patterns.
Second, it is efficient and does not require learning codec parameters: the cosine basis is fixed and data-independent, while the retained coefficient set is selected using a lightweight data-driven tuning procedure.
Third, the orthonormal DCT preserves energy by Parseval's theorem:
$
\lVert x^{3D}\rVert_2^2 = \lVert \mathcal{C}\rVert_2^2.
$
Consequently, the energy preserved by a selected coefficient set is:
\[
\operatorname{EE}_g(\mathcal{I}_g)
=
\frac{\sum_{(h,w,t)\in\mathcal{I}_g}\mathcal{C}_{hwt}^{2}}
{\sum_{h,w,t}\mathcal{C}_{hwt}^{2}}
\in[0,1].
\]

\paragraph{Choosing the embedding dimension.}
We tune the retained coefficients independently for each signal using sampled training windows.
For every coefficient index $(h,w,t)$, we aggregate its energy across the sampled windows and rank coefficients from most to least informative.
We then retain the smallest prefix of this ranking that reaches a target explained-energy threshold, subject to a maximum coefficient budget $K_{\max}$.
This ranked strategy preserves the compactness of DCT3D while avoiding the rigid low-pass behavior of a fixed truncation block.

\subsubsection{\textbf{VAE}}
\label{subsubsec:vae}
As an alternative to our default fixed-basis DCT3D embeddings, we also evaluate learned embeddings obtained by using Variational Autoencoders (VAEs)~\cite{kingma2013auto,prince2023understanding} trained for the signals required by the \bench\ tasks from \textit{Group~1} and \textit{Group~2}.

In our framework, a VAE instantiates a signal-specific codec: the VAE encoder $E_g$ maps a window/chunk $x$ of signal $g$ to a latent-space representation $z \in \mathbb{R}^{K_g}$.
The latent dimensionality is specified per signal by the VAE configuration, providing a compact representation whose compression ratio varies across diagnostics.

The \emph{codec encoder} architecture is chosen according to the signal geometry, using one- or two-dimensional convolutional layers for structured signals and fully-connected layers for low-dimensional time-series signals.
The \emph{codec decoder} mirrors the codec encoder with reversed dimensionality, mapping latent vectors $z$ back to reconstructions in the input space.
VAEs are trained with the standard reconstruction-plus-regularization objective, and at inference time we use the codec-encoder output as the token embedding.

\section{\mmt{} Model}
\label{sec:model}

After the tokenization step is done, \mmt{} maps a variable-length set of token embeddings $\{z_i\}_{i=1}^{L}$ to predictions for a set of output targets. An overview of the model architecture is shown in Fig.~\ref{fig:architecture}.
The model is organized into three components: (i) a \emph{Token Encoder}, that maps each token embedding into a shared $d$-dimensional space and augments it with learned metadata lookup vectors; (ii) a \emph{Transformer Processor}, that produces contextualized token representations; and (iii) an \emph{Output Decoder}, that maps the pooled representation into target predictions.
The Output Decoder is further decomposed into two blocks: modality-specific \textit{MLP heads} and per-target \textit{Output Adapters}, which provide the flexibility to support different output schemas across tasks.

\subsection{Token Encoder}
\label{sec:token_encoder}

Each token corresponds to one embedded input/actuator chunk and consists of a content embedding $z_i\in\mathbb{R}^{K_{g(i)}}$, $i\in\{1,\dots,L\}$ together with discrete metadata: signal ID $g(i)$, modality ID $\mu(i)$, role ID $r(i)$, and a relative position index $p(i)$.
Let $d$ denote the Transformer model dimension. The Token Encoder maps each token to a shared $d$-dimensional representation by (i) projecting the embedding vector with a signal-specific linear map, and (ii) adding learned lookup vectors for the discrete metadata as

\[
h_i^{(0)} = W_{g(i)} z_i
+ e^{\text{sig}}_{g(i)}
+ e^{\text{mod}}_{\mu(i)}
+ e^{\text{role}}_{r(i)}
+ e^{\text{pos}}_{p(i)}
\in \mathbb{R}^{d},
\]
where $W_{g(i)}$ projects $z_i$ into $\mathbb{R}^d$ and $e^{(\cdot)}$, learnable $d$-dimensional vectors indexed by the token metadata.
This factorized representation supports schema changes by composing signal identity, modality, role, and temporal position.

Intuitively, $e^{\text{sig}}_{g(i)}$ conditions the model on signal identity (and is used to define availability masks), $e^{\text{mod}}_{\mu(i)}$ captures modality-specific structure, $e^{\text{role}}_{r(i)}$ distinguishes sensor inputs from actuators, and $e^{\text{pos}}_{p(i)}$ encodes position/recency.
Relative positions are defined within each role by counting backwards from the newest retained chunk: the most recent chunk has $p=1$, the next has $p=2$, and so on.
This provides a task-agnostic notion of recency and supports variable context lengths by changing only the number of retained chunks/tokens, without modifying the backbone.

We prepend a learnable \texttt{[CLS]} (classification) token embedding $h_{\texttt{cls}}^{(0)}\in\mathbb{R}^{d}$ and form the initial transformer input as
\[
H^{(0)}=\big[h_{\texttt{cls}}^{(0)}, h_1^{(0)}, \dots, h_L^{(0)}\big]\in\mathbb{R}^{(L+1)\times d}
\]
together with an attention mask that excludes padded or missing tokens.
The final \texttt{[CLS]} embedding serves as a pooled window representation used by the Output Decoder.

\subsection{Transformer Backbone}
\label{sec:backbone}
The Transformer processor follows a standard Transformer encoder architecture \cite{vaswani2017attention} that applies masked self-attention over the token sequence.
It consists of $N_{\text{layers}}$ stacked transformer blocks with multi-head self-attention and position-wise feed-forward layers, using residual connections and normalization.
Self-attention is masked to exclude padded or missing tokens, enabling variable-length token sets.

Given the token-encoder output $H^{(0)}\in\mathbb{R}^{(L+1)\times d}$, each layer produces
\[
H^{(\ell+1)} = \mathrm{TransformerEnc}^{(\ell)}\!\Big(H^{(\ell)}\Big),
\qquad \ell=0,\dots,N_{\text{layers}}-1,
\]
yielding contextualized representations $H^{(N_{\text{layers}})}$.
We denote the final \texttt{[CLS]} embedding by
$c = H^{(N_{\text{layers}})}_{\texttt{cls}}\in\mathbb{R}^{d}$, which serves as a pooled window representation for the Output Decoder.

\subsection{Output Decoder}
\label{sec:decoder}
The Output Decoder maps the pooled window representation to predictions for each output target.
It is structured in two stages: modality-specific MLP heads that produce latent representations, followed by per-target Output Adapters that map to the appropriate output embedding space.

\paragraph{\textbf{Modality Heads.}}
We obtain modality-specific latent representations from the pooled embedding $c$.
For each modality $\mu$ (e.g., time-series, profile, video), a lightweight MLP head $\phi_{\mu}$ maps $c$ to a modality latent as
\[
u_{\mu} = \phi_{\mu}(c) \in \mathbb{R}^{d_{\mu}},
\]
with $d_{\mu}=d$ in our experiments.

\paragraph{\textbf{Output Adapters.}}
Each target signal $o\in O$ is predicted in its embedding space via an Output Adapter that maps from the corresponding modality latent to the target embedding dimension $K_o$ as
\[
\hat{y}_{o} = \psi_{o}\!\left(u_{\mu(o)}\right) \in \mathbb{R}^{K_o},
\]
where $\mu(o)$ denotes the modality of output $o$, and $\psi_o$ is a small linear/MLP adapter.

\paragraph{\textbf{Outputs.}}
The model produces predictions $\{\hat{y}_{o}\}_{o\in O}$ in the data embedding space.
During evaluation, predictions are mapped back to native signal space using the corresponding decoding operator (e.g., IDCT for DCT3D, or a learned decoder for VAEs).
Training losses and adaptation strategies (including warm-start and freezing policies) are described in the next section.

\section{\mmt{} Training and Adaptation}
\label{sec:training_adaptation}
This section describes how \mmt{} is trained and how a pretrained model is adapted to new tasks and output schemas.
We define the masked objective, summarize how missing inputs/targets are handled, and present our warm-start and freezing strategies.

\subsection{Training Objective and Masking}
\label{sec:loss_embedding_space}
\mmt{} supports two complementary training objectives. Emb\-edding-space supervision is computationally efficient for high-dimensional outputs, while native-space supervision is useful when sparse target observations must be preserved.

\paragraph{Embedding-space loss.}
For each output signal $o\in O$, the model predicts an embedding vector $\hat{y}_o \in \mathbb{R}^{K_o}$.
Preprocessing provides the corresponding target embedding $y_o \in \mathbb{R}^{K_o}$ together with an availability mask $m_o\in\{0,1\}$ indicating whether $o$ is present for the current window.
We define the masked embedding-space mean-squared error as
\begin{equation}
\label{eq:masked_mse}
\mathcal{L}_{\mathrm{emb}}
= \sum_{o\in O} \lambda_o \, m_o \,\frac{1}{K_o}\left\lVert \hat{y}_o - y_o \right\rVert_2^2,
\end{equation}
where $\lambda_o$ are optional per-output weights, and targets with $m_o=0$ do not contribute to the loss.
This objective provides a uniform supervision interface across modalities and reduces the dimensionality of high-rate signals, although achievable native-space accuracy is bounded by the embedding reconstruction error.

\paragraph{Native-space sparse loss.}
For sparse targets, predictions can instead be decoded differentiably and compared against the original non-imputed native targets.
Let $D_o$ denote the decoder associated with output $o$, $\hat{x}_o=D_o(\hat{y}_o)$ the decoded prediction, and $v_{o,p}\in\{0,1\}$ whether native target position $p$ is observed.
We define the native-space sparse loss as
\begin{equation}
\label{eq:native_sparse_mse}
\mathcal{L}_{\mathrm{native}}
= \sum_{o\in O} \lambda_o \, m_o \,
\frac{\sum_{p} v_{o,p}\left(\hat{x}_{o,p}-x_{o,p}\right)^2}
{\sum_{p} v_{o,p}},
\end{equation}
where outputs without observed native positions do not contribute to the loss.
This objective prevents missing target values from becoming artificial supervision.

\paragraph{Masking.}
The masking mechanisms have distinct roles: the attention mask excludes unavailable or padded input tokens from self-attention; $m_o$ excludes entirely unavailable outputs; and $v_{o,p}$ excludes missing positions within an otherwise available native output.
Optionally, training can be regularized by randomly dropping a subset of input/actuator tokens (or entire time chunks) and masking a subset of targets.
As a result, \mmt{} can be trained and evaluated on arbitrary subsets of available input modalities and target signals without requiring a fixed schema.

\subsection{Warm-Start and Task Adaptation}
\label{sec:warmstart_adaptation}

We follow a two-stage protocol. First, we \emph{pretrain} a single model on a broad mixture of signals/tasks in \bench{} to learn a general-purpose representation for tokamak plasma dynamics. Second, for each downstream task, we \emph{fine-tune} a subset of model blocks together with task-specific Output Adapters (Section~\ref{sec:experiments}).

To adapt efficiently under task and schema variations (changes in $\mathcal{I}$, $\mathcal{A}$, $\mathcal{O}$, or embedding dimensions), \mmt{} supports warm-starting and (optionally) freezing the main model blocks, namely Token Encoder, Transformer Backbone, Modality Heads, and Output Adapters (Fig.~\ref{fig:architecture}).
Warm-start uses partial loading: parameters are reused only when both their identifier and tensor shape match; mismatched or missing parameters remain at initialization.
Consequently, common adaptation scenarios are handled predictably: adding a new input signal introduces a new projection layer in the token encoder while reusing the backbone; adding a new output introduces a new Output Adapter; changing an embedding dimension reinitializes only the affected projections/adapters; and changing the number of retained chunks affects sequence length at runtime without altering backbone parameters.
Table~\ref{tab:flexibility} summarizes supported scenarios of changes and the resulting model behavior.

%
\begin{table}[t]
\centering
\caption{List of possible changes in the inputs and outputs between model runs and the corresponding adaptations.
}
\small
\begin{tabular}{>{\raggedright\arraybackslash}p{0.39\linewidth} p{0.01\linewidth} p{0.47\linewidth}}
\toprule
\textbf{Change between runs} & \textbf{} & \textbf{Model adaptation} \\
\midrule
Remove some inputs / actuators & \cmark & Fewer tokens; Transformer Backbone unchanged \\
Add new input / actuator & \cmark & New Token Encoder projection initialized; other blocks reused \\
Remove outputs & \cmark & Corresponding Output Adapters absent; loss ignores them via masks \\
Add new outputs & \cmark & New Output Adapters initialized; existing adapters reused \\
Change embedding dim / encoder for an existing signal & \cmark & Token Encoder / Output Adapter reinitialized (shape mismatch); other blocks reused \\
Change chunk length / max\_chunks / stride & \cmark & Different token count; relative-position encoding preserves a consistent notion of recency \\
Evaluate on a subset of inputs/outputs & \cmark & Same masking mechanism; evaluation works on arbitrary subsets \\
Change $d$ (model dimension) / Backbone shape & \xmark & Requires training a new Backbone \\
\bottomrule
\end{tabular}
\label{tab:flexibility}
\end{table}
%

\section{Experiments}
\label{sec:experiments}

\subsection{Experimental Protocol}
\label{subsec:exp_protocol}

We evaluate two downstream regimes: task-specific \emph{fine-tuning} of a pretrained \ac{fm}, and task-specific \emph{training from scratch} under the same epoch budget.
We first pretrain \mmt{} on a broad multi-signal objective to obtain the initialization used for fine-tuning.

\paragraph{\ac{fm} pretraining.}
We pretrain \mmt{} on a broad signal set defined in our pretraining configuration.
This pretrained checkpoint defines the \textit{initialization reused for all downstream fine-tuning experiments.}
Inputs include \emph{all benchmark input diagnostics} and \emph{all actuators} (16 sensor inputs and 5 actuators).
Targets include \emph{all benchmark inputs} and \emph{all benchmark outputs} (34 targets).
Pretraining uses a reconstruction objective on 50\,ms windows.
This broad multi-signal objective encourages the backbone to learn a general representation that transfers across tasks.
We pretrain two model sizes, \emph{Base} (\(6.93\)M parameters) and \emph{Tiny} (\(3.69\)M parameters), to study the effect of model capacity.
Notably, both variants remain lightweight, supporting efficient training and evaluation in this setting.

\paragraph{Task-specific fine-tuning (warm-start).}
For each downstream task, we initialize from the pretrained checkpoint and fine-tune in two stages.
We warm-start the Token Encoder, Transformer Backbone, and Modality Heads; Output Adapters are instantiated per task since output sets and embedding dimensions are task-dependent.
Stage~1 (\emph{task-facing adaptation}) freezes the \emph{Token Encoder} and \emph{Transformer Backbone}, updating only the Modality Heads and Output Adapters.
Stage~2 (\emph{end-to-end refinement}) unfreezes all components and jointly refines the full model.
This staged protocol first adapts the task-facing components while preserving pretrained representations, then allows end-to-end refinement.

\paragraph{Training from scratch.}
To isolate the benefits of pretraining, we also train task-specific \mmt{} models from random initialization.
These models use the same epoch budget as the fine-tuned models and are trained independently for the \emph{14} benchmark tasks.

\paragraph{Embedding study.}
We additionally perform a small embedding study on Group~1 and Group~2 tasks, comparing our default DCT3D embedding to a VAE-based embedding under the same downstream training protocol.

\paragraph{Hyperparameters.}
Exact model hyperparameters, optimization settings, training schedules, and computational resources are reported in Appendix~\ref{app:hyper-param}.

\subsection{Evaluation Protocol}
\label{sec:eval_protocol}


We evaluate all models on the \bench{} benchmark, following its task definitions, data splits, and sparse-evaluation protocol.
We compare against four \bench{} baselines: two statistical predictors (\emph{Mean} and \emph{Persistence}), and two neural models, namely, \emph{CNN} and \emph{CNN+LSTM}. We report test-set performance using the benchmark’s normalized RMSE (NRMSE) aggregated per task and per group.
Our models predict targets in embedding space; for evaluation, predictions are decoded to native signal space using the embedding-specific decoding operator.
Metrics are computed against the original non-imputed native targets and only at observed target positions, so missing target values do not contribute to NRMSE.
We refer the reader to the \bench{} paper for the formal metric definitions and aggregation procedure.

\subsection{Results}
\label{sec:results}


\begin{table}[t]
\centering
\caption{Test NRMSE on \bench{}. \textbf{C+LSTM} denotes the CNN+LSTM baseline and \textbf{Pers.} the Persistence baseline. Dashes indicate tasks for which Persistence is not defined. Best and second-best values per row are shown in bold and underlined, respectively. Group and overall averages are unweighted means over tasks.}
\small
\setlength{\tabcolsep}{1.8pt}
\renewcommand{\arraystretch}{0.95}
\begin{tabular}{lrrrrrrr}
\toprule
& \multicolumn{4}{c}{\textbf{Baselines}} & \multicolumn{3}{c}{\textbf{\mmt}} \\
\cmidrule(lr){2-5}\cmidrule(lr){6-8}
\textbf{Task} & \textbf{Mean} & \textbf{Pers.} & \textbf{CNN} & \textbf{C+LSTM} & \textbf{Scratch} & \textbf{FT-Tiny} & \textbf{FT-Base} \\
\midrule
1-1 & 0.757 & --- & 0.209 & 0.222 & \textbf{0.140} & 0.153 & \underline{0.150} \\
1-2 & 0.998 & --- & 0.047 & 0.051 & \textbf{0.042} & 0.044 & \underline{0.043} \\
1-3 & 0.927 & --- & 0.148 & 0.159 & \underline{0.129} & \textbf{0.122} & \textbf{0.122} \\
\midrule
\emph{Group 1} & 0.894 & --- & 0.135 & 0.144 & \textbf{0.104} & 0.106 & \underline{0.105} \\
\midrule
2-1 & 0.817 & --- & 0.275 & 0.220 & \textbf{0.180} & 0.192 & \underline{0.181} \\
2-2 & 0.991 & --- & 0.066 & 0.057 & \textbf{0.042} & 0.045 & \underline{0.043} \\
2-3 & 0.923 & --- & 0.136 & 0.126 & \underline{0.094} & 0.097 & \textbf{0.092} \\
\midrule
\emph{Group 2} & 0.911 & --- & 0.159 & 0.134 & \textbf{0.105} & \underline{0.111} & \textbf{0.105} \\
\midrule
3-1 & 0.984 & 0.536 & 0.396 & 0.374 & \textbf{0.335} & 0.355 & \underline{0.346} \\
3-2 & 0.630 & 0.365 & 0.331 & 0.335 & \textbf{0.265} & 0.279 & \underline{0.269} \\
3-3 & 0.841 & --- & 0.386 & 0.319 & \textbf{0.217} & 0.231 & \underline{0.220} \\
\midrule
\emph{Group 3} & 0.818 & --- & 0.371 & 0.343 & \textbf{0.273} & 0.288 & \underline{0.278} \\
\midrule
4-1 & 0.511 & 0.325 & 0.275 & 0.227 & 0.266 & \underline{0.223} & \textbf{0.213} \\
4-2 & 0.511 & 0.325 & 0.274 & \underline{0.229} & 0.275 & 0.230 & \textbf{0.214} \\
4-3 & 0.731 & --- & 0.129 & 0.118 & 0.141 & \underline{0.095} & \textbf{0.080} \\
4-4 & 1.027 & 0.718 & 0.530 & 0.404 & 0.420 & \underline{0.384} & \textbf{0.349} \\
4-5 & 0.653 & 0.870 & 0.651 & \textbf{0.649} & 0.661 & \underline{0.650} & 0.651 \\
\midrule
\emph{Group 4} & 0.687 & --- & 0.372 & 0.326 & 0.353 & \underline{0.317} & \textbf{0.301} \\
\midrule
\textit{Overall} & 0.807 & --- & 0.275 & 0.249 & 0.229 & \underline{0.221} & \textbf{0.212} \\
\bottomrule
\end{tabular}
\label{tab:main_results}
\end{table}

\begin{figure}
    \includegraphics[width=0.45\textwidth]{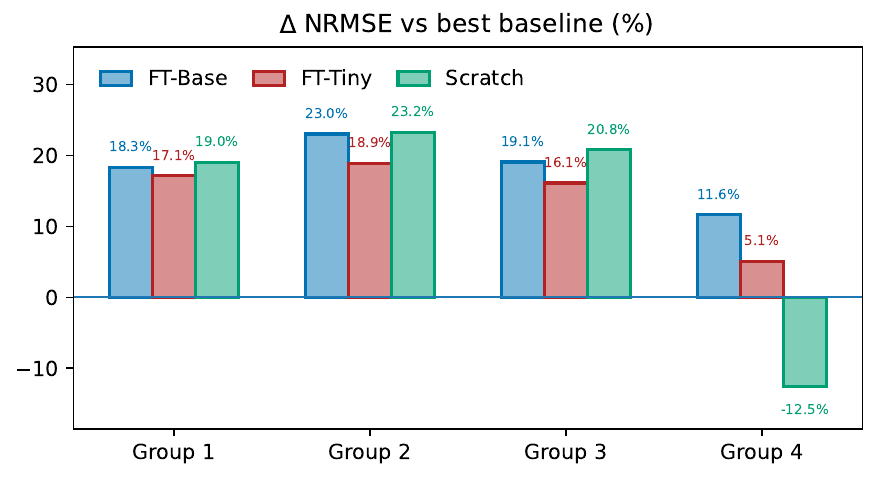}
    \captionsetup{skip=2pt}
    \caption{Task-level relative NRMSE improvement over the strongest available benchmark baseline, averaged within each group. Positive values indicate lower NRMSE; negative values indicate worse performance.
    \vspace{-0.3cm}}
    \label{fig:per_group_improvement}
\end{figure}



Table~\ref{tab:main_results} reports per-task test NRMSE for the benchmark baselines and the three \mmt{} variants: \emph{FT-Base}, \emph{FT-Tiny}, and \emph{Scratch}. Figure~\ref{fig:per_group_improvement} summarizes task-level relative improvements over the strongest available benchmark baseline, averaged within each group.
Appendix Table~\ref{tab:per_signal_all_groups} reports signal-level results together with the corresponding DCT3D compression statistics for each output.

\paragraph{Main findings.} The major comparative results are listed below:
\begin{itemize}[leftmargin=2em]

    \item \textbf{Consistent gains over the benchmark baselines:}
    \emph{FT-Base} outperforms the best benchmark baseline on all tasks except Task~4-5 and across all four benchmark groups. \emph{FT-Tiny} also remains competitive, outperforming the best benchmark baseline on all tasks except Tasks~4-2 and 4-5.

    \item \textbf{Tiny model retains most of the performance:}
    \emph{FT-Tiny} closely tracks \emph{FT-Base} across groups with only minor degradations, suggesting that much of the benefit is retained at substantially lower capacity.

    \item \textbf{Pretraining helps most on harder tasks:}
    In Group~4, \emph{FT-Base} outperforms \emph{Scratch} on all five tasks. As shown in Fig.~\ref{fig:per_group_improvement}, the fine-tuned variants improve over the strongest benchmark baseline at group level, whereas training from scratch does not. Warm-starting also improves the high-dimensional equilibrium reconstruction and forecasting tasks (Tasks~1-3 and 2-3). Together, these results suggest that pretraining is especially useful for complex outputs and longer prediction horizons.

\end{itemize}

\paragraph{Task~4-5.}
Task~4-5 differs from the other forecasting tasks because its targets are spectral features of Mirnov signals rather than time-domain waveforms.
Following \bench{}, we evaluate this task in the Fourier domain.
Performance remains close across the benchmark baselines and \mmt{} variants. Since the targets are already represented in the Fourier domain, applying an additional DCT3D embedding does not provide a clear advantage under the current formulation.

\paragraph{Signal-level metrics.}
Appendix Table~\ref{tab:per_signal_all_groups} provides a finer-grained view of performance across individual output signals.
\emph{FT-Base} outperforms the best benchmark baseline on 52 of 57 outputs, showing that the aggregate improvements are broadly distributed rather than driven by a small subset of targets.
DCT3D acts as an identity or near-identity mapping for low-dimensional equilibrium scalars, while substantially compressing higher-dimensional outputs.
The largest coefficient budgets are required by high-frequency and long-horizon targets, particularly in Tasks~3-2, 4-1, 4-2, and 4-5.



\begin{table}[t]
\centering
\caption{Group~1 and Group~2 test NRMSE for VAE-based embeddings under FT-Base and Scratch regimes. $\Delta$ reports the NRMSE difference relative to the corresponding DCT3D result under the same training regime.}
\small
\setlength{\tabcolsep}{3.0pt}
\renewcommand{\arraystretch}{0.95}
\begin{tabular}{lcccc}
\toprule
& \multicolumn{2}{c}{\textbf{VAE FT-Base}} & \multicolumn{2}{c}{\textbf{VAE Scratch}} \\
\cmidrule(lr){2-3}\cmidrule(lr){4-5}
\textbf{Task} & \textbf{NRMSE} & $\Delta$ & \textbf{NRMSE} & $\Delta$ \\
\midrule
1-1 & 0.165 & +0.015 & 0.153 & +0.013 \\
1-2 & 0.041 & -0.002 & 0.046 & +0.004 \\
1-3 & 0.122 & +0.000 & 0.130 & +0.001 \\
\midrule
\emph{Group 1} & 0.109 & +0.004 & 0.110 & +0.006 \\
\midrule
2-1 & 0.198 & +0.017 & 0.189 & +0.009 \\
2-2 & 0.045 & +0.002 & 0.043 & +0.001 \\
2-3 & 0.099 & +0.007 & 0.097 & +0.003 \\
\midrule
\emph{Group 2} & 0.114 & +0.009 & 0.110 & +0.005 \\
\bottomrule
\end{tabular}
\label{tab:group1_2_vae}
\end{table}


\paragraph{Embedding study: DCT3D vs VAE}
We conduct a focused embedding study on Group~1 and Group~2 to compare our default fixed-basis DCT3D codecs against a learned VAE codecs.
VAE codecs are pretrained on MAST data for the corresponding Group~1 and Group~2 signals.
%
We compare DCT3D and VAE under the same downstream training protocol, considering both training from scratch and task-specific fine-tuning.
Since fine-tuned models are warm-started from a DCT3D-pretrained checkpoint, this initialization may be suboptimal for VAE-based embeddings; therefore, training from scratch provides the fairest assessment of the embedding choice.

VAE results for Group~1 and Group~2, together with their $\Delta$ relative to the corresponding DCT3D models, are reported in Table~\ref{tab:group1_2_vae}.
Appendix Table~\ref{tab:group1_2_inputs_dct3d_vae} reports the corresponding Group~1 and Group~2 input and actuator shapes and DCT3D/VAE embedding dimensions.
Overall, DCT3D remains slightly stronger than the VAE variants in this setting, indicating that a simple frequency-based representation is already competitive for these signals and horizons.
At the same time, the VAE variants achieve close performance to DCT3D and still outperform the best benchmark baseline reported in Table~\ref{tab:main_results}.
Moreover, Appendix Table~\ref{tab:group1_2_inputs_dct3d_vae} shows that VAE embeddings often provide higher compression ratios than DCT3D for the Group~1 and Group~2 inputs and actuators.
These results suggest that learned codecs are a promising direction, but also that DCT3D is a robust and flexible default for schema-flexible task adaptation.
We expect further gains from a dedicated learned-embedding study (e.g., VAE-specific pretraining initializations, objectives, and hyperparameter tuning), which we leave to future work.

\section{Limitations and Ethical Considerations}
\label{sec:limitations}

Our evaluation is limited to the publicly available MAST dataset and does not establish cross-device generalization.
While warm-start fine-tuning provides evidence of transfer across related tasks, evaluating pretrained representations on other tokamaks
and operating regimes remains an important next step. The comparison is also constrained by the availability of benchmark-compatible baselines,
and our VAE study is not an exhaustive exploration of learned codecs. Rapidly varying signals and long-horizon forecasting remain challenging
settings, motivating further work on frequency-aware objectives, alternative chunking strategies, and regime-wise error analysis.

The MAST data contain experimental plasma diagnostics and no personal data. Nevertheless, prediction errors could have significant consequences
if models were used for plasma monitoring or control. \mmt{} is intended as a research framework and should not be deployed
in safety-critical settings without device-specific validation, uncertainty quantification, robustness testing under distribution
shift and missing diagnostics, and appropriate human oversight.

\section{Conclusions}
\label{sec:conclusion}

\mmt{} is a modular, multi-modal transformer designed for tokamak data characterized by heterogeneous modalities, multi-rate sampling, and partial observability.
\mmt{} tokenizes windowed diagnostics and actuators into a variable-length set of token embeddings, and processes them with a shared Transformer Backbone.
A two-stage Output Decoder composed of modality-specific Heads and per-target Output Adapters enables flexible prediction across tasks with heterogeneous objectives and output schemas, while explicit masks and configurable local imputation support missing inputs and targets.

On the \bench{} benchmark, \emph{FT-Base} improves over the best benchmark baseline on all tasks except Task~4-5 and across all groups.
Although warm-starting does not uniformly outperform training from scratch, its benefits are clearest on the most demanding downstream tasks: \emph{FT-Base} improves over \emph{Scratch} on all Group~4 tasks and on the high-dimensional equilibrium reconstruction and forecasting tasks.
These results support the foundation-model perspective: broad multi-signal pretraining yields a transferable initialization that is particularly valuable for complex outputs and longer prediction horizons.
A pretrained, schema-flexible model can therefore provide a reusable starting point for data-driven tokamak models across heterogeneous tasks and output schemas.


Future work will focus on extending \mmt{} beyond MAST to other tokamaks and reactor settings, enabling broader cross-device generalization by exploiting its modular design and schema-flexible tokenization.
Another direction is a more extensive study of embedding choices---including learned codecs and task-adaptive representations---which may further improve performance, particularly for high-frequency targets.
We also plan to investigate the integration of pretrained PDE foundation models as physics-aware priors for plasma dynamics, to assess whether such models \cite{herde2024poseidon} can improve data efficiency and long-horizon forecasting when combined with multi-modal fusion.
Beyond plasma fusion, the proposed tokenization framework is applicable to many multi-channel scientific problems with heterogeneous sensors, missing channels, and varying schemas.


\begin{acks}
This work was supported by the Hartree National Centre for Digital Innovation, a collaboration between STFC, IBM, and UKAEA. We also
wish to extend our gratitude to the MAST Team for their efforts in collecting the diagnostic source data during the operation of the MAST device.
\end{acks}


\bibliographystyle{ACM-Reference-Format}
\bibliography{references}

%
%

\appendix

\renewcommand{\thetable}{A.\arabic{table}}
\setcounter{table}{0}

\section{Hyperparameters and Implementation Details}
\label{app:hyper-param}

\subsection{DCT3D tuning}
\label{app:dct3d_tuning}

We tune DCT3D independently for each signal using ranked coefficient selection (Section~\ref{subsubsec:dct3d}). Each run samples windows from 100 training shots (at most 15{,}000 windows), ranks coefficients by aggregate energy, and retains the smallest prefix reaching the role-specific explained-energy threshold under a hard budget. We enforce minimum index coverage: 5 temporal indices for time series; 10 channel and 5 temporal indices for profiles; and 10 indices per spatial dimension plus 5 temporal indices for videos.

Table~\ref{tab:group1_2_inputs_dct3d_vae} reports the source codec settings for Group~1 and Group~2 inputs and actuators, while
Table~\ref{tab:per_signal_all_groups} reports the tuned DCT3D settings used for all outputs across tasks.

For pretraining, we tune the encoders using explained-energy thresholds of 0.999 for inputs and actuators, and 0.995 for outputs, with a maximum budget of $K_{\max}=4{,}096$.

For downstream task adaptation, warm-start runs reuse the pretrained input and actuator codecs to preserve alignment with the source model, while task-specific output codecs are retuned using a stricter explained-energy threshold of 0.999 and a maximum budget of $K_{\max}=8{,}192$ coefficients.
For Scratch runs, all codecs are tuned independently using the same downstream policy.
The larger downstream budget is reached only by a small subset of high-dimensional or long-horizon outputs, namely Tasks~3-2, 4-1, 4-2, and 4-5.

\subsection{Model Architectures}
\label{app:model_params}

We report two pretrained model sizes: \emph{Base} (6{,}927{,}799 parameters) and \emph{Tiny} (3{,}694{,}391 parameters).
Both use the same modular structure described in Section~\ref{sec:model}, with them differing only in Transformer Backbone capacity (e.g., $d$, number of layers/heads, FF dimension). \\

For the \emph{Base} model, the Transformer Backbone uses $d=192$, 4 layers, 6 heads, FF dimension 768, and dropout 0.05, with GELU activation.
For the \emph{Tiny} model, the Transformer Backbone uses $d=128$, 2 layers, 4 heads, FF dimension 384, and dropout 0.05, with GELU activation.
Modality Heads use per-modality MLPs with hidden dimension and output dimension $d$.

\paragraph{Output Adapter policy.}
Pretraining uses linear Output Adapters, while downstream fine-tuned/scratch runs enable bucketed hidden dimensions for Output Adapters based on output embedding size.
We use the following bucket rules in downstream runs: outputs with embedding dimension $\le 64$ use hidden size $0$, $\le 512$ use $32$, $\le 4096$ use $64$, $\le 8192$ use $128$, and larger outputs use hidden size $d$.

\subsection{Training Hyperparameters}
\label{app:train_params}

All models are trained with \textit{AdamW} and automatic mixed precision (AMP).
We use early stopping with patience 10 and $\Delta=0$.
Learning-rate schedules use a warmup fraction of 0.02.

\paragraph{Pretraining.}
Pretraining is run for 50 epochs with batch size 512.
We use per-block learning rates: Token Encoder $5\times10^{-3}$, Transformer Backbone $10^{-3}$, Modality Heads $5\times10^{-3}$, and Output Adapters $5\times10^{-3}$. Learning-rate schedules use a linear warmup over the first 2\% of training steps, followed by cosine decay to zero over the remaining steps.

\paragraph{Fine-tuning (two stages).}
Fine-tuning proceeds in two stages: 5 epochs with the Token Encoder and Transformer Backbone frozen, followed by 15 epochs of end-to-end refinement with all components unfrozen.
Stage-specific learning rates are:
\begin{itemize}[leftmargin=2em]
  \item \emph{Stage 1 (task-facing adaptation):} Modality Heads $10^{-3}$, Output Adapters $5\times10^{-3}$.
  \item \emph{Stage 2 (end-to-end refinement):} Token Encoder $5\times10^{-4}$, Transformer Backbone $5\times10^{-4}$, Modality Heads $10^{-3}$, Output Adap\-ters $5\times10^{-3}$.
\end{itemize}
Weight decay is 0.01 on the trainable blocks in each stage.
Fine-tuning uses a batch size of 512 windows for all tasks, except Group~4 tasks where we use 256 due to higher memory requirements.

\paragraph{Missing-value policy.}
During pretraining, we retain partially observed signals and zero-fill remaining non-finite values before DCT3D tuning and encoding. Since signals are standardized before embedding, zero filling corresponds to mean imputation. During downstream task adaptation, we retain partially observed signals and instead apply temporal interpolation followed by spatial interpolation and a zero fallback on local copies, preserving the original native targets for sparse supervision and evaluation.
This mixed policy uses a simple and stable fallback during broad multi-signal pretraining, while providing smoother task-specific inputs during downstream adaptation.

\paragraph{Loss selection.}
We use the native-space sparse loss in Eq.~\eqref{eq:native_sparse_mse} by default because evaluation is performed in native signal space with missing-value masking. This aligns supervision with the benchmark metric. For Tasks~3-2, 4-1, 4-2, and 4-5, we instead use embedding-space supervision because their tuned DCT3D output representations remain high-dimensional or reach the $K_{\max}=8{,}192$ coefficient cap. In these cases, native decoding is more expensive and direct regression in the capped embedding space is empirically more stable.

\paragraph{Windowing, chunking, and collation.}
All experiments use nonoverlapping 5\,ms chunks. For non-Markovian tasks, sequence lengths cover 100\,ms of past context plus the forecast interval: up to 21 chunks per role for Task~3-3, with a 5\,ms horizon, and 40 for Group~4, with a 100\,ms horizon. To reduce redundancy, candidate windows are subsampled with strides of 10\,ms during pretraining, 5\,ms during downstream adaptation for Groups~1--3, and 20\,ms for Group~4. Batches are padded to the maximum token length with attention masks. During collation, input signals and chunks are randomly dropped with probability 0.10 in pretraining and 0.05 in downstream adaptation to improve robustness to missing diagnostics.

\paragraph{Training from scratch.}
Scratch runs follow the same architecture as the Base fine-tuned model, and use the same epoch budget as fine-tuning (20 epochs).

\begin{table*}[b!]
\vspace{2em}
\centering
\caption{
Group~1 and Group~2 input and actuator signals. $(H,W,T)$ denotes the native tensor shape for a single 5\,ms chunk, with $\mathrm{dim}=HWT$. DCT3D columns report the source codecs used by warm-start models: embedding dimension $K_g$, compression ratio $\mathrm{dim}/K_g$, and explained energy (EE). VAE columns report the latent embedding dimension $K_g$ and the corresponding compression ratio. Inputs are shared by Groups~1 and~2, while actuators are used only by Group~2.
}
\small
\setlength{\tabcolsep}{4pt}
\renewcommand{\arraystretch}{1.05}
\scalebox{0.92}{
\begin{tabular}{l r r r r r r}
\toprule
\textbf{Signal} &
\textbf{(H, W, T)} &
\multicolumn{3}{c}{\textbf{DCT3D}} &
\multicolumn{2}{c}{\textbf{VAE}} \\
\cmidrule(lr){3-5}\cmidrule(lr){6-7}
 &  & $\bm{K_g}$ & \textbf{dim/$K_g$} & \textbf{EE} & $\bm{K_g}$ & \textbf{dim/$K_g$}\\
\midrule
\multicolumn{7}{l}{\textit{\textbf{Inputs shared by Groups~1 and~2}}} \\
magnetics-flux\_loop\_flux & $(15, 1, 25)$ & 44 & 8.52 & 1.000 & 12 & 31.25 \\
magnetics-b\_field\_pol\_probe\_ccbv\_field & $(40, 1, 25)$ & 54 & 18.52 & 0.999 & 30 & 33.33 \\
magnetics-b\_field\_pol\_probe\_obr\_field & $(18, 1, 25)$ & 57 & 7.89 & 1.000 & 15 & 30.00 \\
magnetics-b\_field\_pol\_probe\_obv\_field & $(18, 1, 25)$ & 45 & 10.00 & 1.000 & 15 & 30.00 \\
magnetics-b\_field\_tor\_probe\_saddle\_voltage & $(12, 1, 250)$ & 28 & 107.14 & 1.000 & 8 & 375.00 \\
pf\_active-coil\_current & $(10, 1, 20)$ & 26 & 7.69 & 1.000 & 6 & 33.33 \\
pf\_active-solenoid\_current & $(1, 1, 20)$ & 5 & 4.00 & 1.000 & 4 & 5.00 \\
summary-ip & $(1, 1, 20)$ & 5 & 4.00 & 1.000 & 3 & 6.67 \\
\midrule
\multicolumn{7}{l}{\textit{\textbf{Actuators used by Group~2 only}}} \\
pf\_active-coil\_voltage & $(4, 1, 20)$ & 22 & 3.64 & 0.999 & 4 & 20.00 \\
summary-power\_nbi & $(1, 1, 20)$ & 5 & 4.00 & 0.999 & 3 & 6.67 \\
\bottomrule
\end{tabular}
}
\label{tab:group1_2_inputs_dct3d_vae}
\end{table*}

\subsection{Computational Resources}
\label{app:compute}
All experiments were run on a single GPU node equipped with an NVIDIA A100-SXM4-80GB GPU (80\,GB VRAM) with CUDA 12.8 (driver 570.195.03).
To amortize preprocessing cost, we cache window-level representations in memory; caching uses \texttt{float16} storage and 32 worker processes for cache materialization.


\begin{table*}[tb!]
\centering
\caption{Detailed per-signal test NRMSE in native space (expanded view of Table~\ref{tab:main_results}). DCT3D columns summarize the task-specific output codecs used by the warm-start models: native tensor shape $(H,W,T)$, embedding dimension $K_g$, compression ratio $\mathrm{dim}/K_g$ with $\mathrm{dim}=HWT$, and explained energy (EE; rounded to 3 decimals). Neural-baseline columns report CNN and CNN+LSTM performance. Best and second-best NRMSE values per row are shown in bold and underlined, respectively.}

\setlength{\tabcolsep}{2.5pt}
\scalebox{0.78}{
\begin{tabular}{c c >{\raggedright\arraybackslash}p{0.35\textwidth}
                r r r r @{\hspace{6pt}} r r @{\hspace{6pt}} r r r}
\toprule
\textbf{Group} & \textbf{Task} & \textbf{Signal} &
\multicolumn{4}{c}{\textbf{DCT3D}} &
\multicolumn{2}{c}{\textbf{Neural Baselines}} &
\multicolumn{3}{c}{\textbf{\mmt}} \\
\cmidrule(lr){4-7}\cmidrule(lr){8-9}\cmidrule(lr){10-12}
 &  &  &
$\bm{(H,W,T)}$ & $\bm{K_g}$ & \textbf{dim/$K_g$} & \textbf{EE} &
\textbf{CNN} & \textbf{C+LSTM} & \textbf{Scratch} & \textbf{FT-Tiny} & \textbf{FT-Base} \\
\midrule
\multirow{18}{*}{\rotatebox[origin=c]{90}{\textbf{Group~1}}} & \multirow{15}{*}{1-1} & equilibrium-beta\_normal & $(1,1,1)$ & 1 & 1.00 & 1.000 & 0.222 & 0.243 & \textbf{0.157} & \underline{0.170} & 0.172 \\
 &  & equilibrium-beta\_pol & $(1,1,1)$ & 1 & 1.00 & 1.000 & 0.224 & 0.230 & \textbf{0.153} & \underline{0.166} & 0.170 \\
 &  & equilibrium-beta\_tor & $(1,1,1)$ & 1 & 1.00 & 1.000 & 0.225 & 0.267 & \textbf{0.157} & 0.175 & \underline{0.173} \\
 &  & equilibrium-bphi\_rmag & $(1,1,1)$ & 1 & 1.00 & 1.000 & 0.285 & 0.286 & \textbf{0.179} & 0.199 & \underline{0.195} \\
 &  & equilibrium-bvac\_rmag & $(1,1,1)$ & 1 & 1.00 & 1.000 & 0.271 & 0.278 & \textbf{0.154} & 0.180 & \underline{0.169} \\
 &  & equilibrium-elongation & $(1,1,1)$ & 1 & 1.00 & 1.000 & 0.232 & 0.236 & \textbf{0.162} & 0.176 & \underline{0.172} \\
 &  & equilibrium-elongation\_axis & $(1,1,1)$ & 1 & 1.00 & 1.000 & 0.235 & 0.239 & \textbf{0.168} & 0.184 & \underline{0.180} \\
 &  & equilibrium-magnetic\_axis\_r & $(1,1,1)$ & 1 & 1.00 & 1.000 & 0.166 & 0.171 & \textbf{0.114} & 0.127 & \underline{0.126} \\
 &  & equilibrium-magnetic\_axis\_z & $(1,1,1)$ & 1 & 1.00 & 1.000 & 0.122 & 0.152 & \textbf{0.075} & 0.078 & \underline{0.075} \\
 &  & equilibrium-minor\_radius & $(1,1,1)$ & 1 & 1.00 & 1.000 & 0.201 & 0.205 & \textbf{0.144} & 0.159 & \underline{0.157} \\
 &  & equilibrium-q95 & $(1,1,1)$ & 1 & 1.00 & 1.000 & 0.162 & 0.198 & \textbf{0.103} & 0.112 & \underline{0.108} \\
 &  & equilibrium-triangularity\_lower & $(1,1,1)$ & 1 & 1.00 & 1.000 & 0.244 & 0.249 & \textbf{0.169} & 0.184 & \underline{0.179} \\
 &  & equilibrium-triangularity\_upper & $(1,1,1)$ & 1 & 1.00 & 1.000 & 0.239 & 0.243 & \textbf{0.169} & 0.183 & \underline{0.179} \\
 &  & equilibrium-x\_point\_r & $(2,1,1)$ & 2 & 1.00 & 1.000 & 0.270 & 0.255 & \textbf{0.170} & 0.177 & \underline{0.174} \\
 &  & equilibrium-x\_point\_z & $(2,1,1)$ & 2 & 1.00 & 1.000 & 0.039 & 0.070 & \textbf{0.022} & 0.024 & \underline{0.022} \\
\cmidrule(lr){2-12}
 & \multirow{2}{*}{1-2} & equilibrium-lcfs\_r & $(170,1,1)$ & 10 & 17.00 & 0.999 & 0.054 & 0.058 & \textbf{0.044} & 0.047 & \underline{0.046} \\
 &  & equilibrium-lcfs\_z & $(170,1,1)$ & 11 & 15.45 & 0.999 & \underline{0.040} & 0.044 & \textbf{0.040} & 0.041 & 0.040 \\
\cmidrule(lr){2-12}
 & \multirow{1}{*}{1-3} & equilibrium-psi & $(65,65,1)$ & 121 & 34.92 & 0.999 & 0.148 & 0.159 & 0.129 & \textbf{0.122} & \underline{0.122} \\
\midrule
\multirow{21}{*}{\rotatebox[origin=c]{90}{\textbf{Group~2}}} & \multirow{18}{*}{2-1} & equilibrium-beta\_normal & $(1,1,5)$ & 5 & 1.00 & 1.000 & 0.359 & 0.309 & \textbf{0.264} & 0.279 & \underline{0.265} \\
 &  & equilibrium-beta\_pol & $(1,1,5)$ & 5 & 1.00 & 1.000 & 0.367 & 0.318 & \textbf{0.262} & 0.278 & \underline{0.265} \\
 &  & equilibrium-beta\_tor & $(1,1,5)$ & 5 & 1.00 & 1.000 & 0.339 & 0.284 & \textbf{0.242} & 0.259 & \underline{0.244} \\
 &  & equilibrium-bphi\_rmag & $(1,1,5)$ & 5 & 1.00 & 1.000 & 0.414 & 0.357 & \textbf{0.280} & 0.299 & \underline{0.281} \\
 &  & equilibrium-bvac\_rmag & $(1,1,5)$ & 5 & 1.00 & 1.000 & 0.324 & 0.271 & \textbf{0.183} & 0.205 & \underline{0.186} \\
 &  & equilibrium-elongation & $(1,1,5)$ & 5 & 1.00 & 1.000 & 0.325 & 0.254 & \textbf{0.214} & 0.229 & \underline{0.219} \\
 &  & equilibrium-elongation\_axis & $(1,1,5)$ & 5 & 1.00 & 1.000 & 0.322 & 0.254 & \underline{0.224} & 0.238 & \textbf{0.224} \\
 &  & equilibrium-magnetic\_axis\_r & $(1,1,5)$ & 5 & 1.00 & 1.000 & 0.202 & 0.170 & \textbf{0.145} & 0.158 & \underline{0.148} \\
 &  & equilibrium-magnetic\_axis\_z & $(1,1,5)$ & 5 & 1.00 & 1.000 & 0.189 & 0.141 & \underline{0.107} & 0.112 & \textbf{0.106} \\
 &  & equilibrium-minor\_radius & $(1,1,5)$ & 5 & 1.00 & 1.000 & 0.253 & 0.211 & \textbf{0.182} & 0.194 & \underline{0.185} \\
 &  & equilibrium-q95 & $(1,1,5)$ & 5 & 1.00 & 1.000 & 0.248 & 0.184 & \underline{0.137} & 0.148 & \textbf{0.137} \\
 &  & equilibrium-triangularity\_lower & $(1,1,5)$ & 5 & 1.00 & 1.000 & 0.305 & 0.251 & \textbf{0.214} & 0.229 & \underline{0.217} \\
 &  & equilibrium-triangularity\_upper & $(1,1,5)$ & 5 & 1.00 & 1.000 & 0.281 & 0.246 & \textbf{0.204} & 0.216 & \underline{0.208} \\
 &  & equilibrium-x\_point\_r & $(2,1,5)$ & 8 & 1.25 & 1.000 & 0.371 & 0.307 & \underline{0.225} & 0.236 & \textbf{0.222} \\
 &  & equilibrium-x\_point\_z & $(2,1,5)$ & 8 & 1.25 & 1.000 & 0.076 & 0.036 & 0.030 & \underline{0.029} & \textbf{0.028} \\
 &  & pf\_active-coil\_current & $(10,1,100)$ & 37 & 27.03 & 0.999 & 0.169 & 0.133 & \textbf{0.082} & \underline{0.085} & 0.087 \\
 &  & pf\_active-solenoid\_current & $(1,1,100)$ & 5 & 20.00 & 1.000 & 0.125 & 0.088 & 0.067 & \underline{0.059} & \textbf{0.055} \\
 &  & summary-ip & $(1,1,100)$ & 8 & 12.50 & 0.999 & 0.282 & \textbf{0.143} & \underline{0.173} & 0.202 & 0.175 \\
\cmidrule(lr){2-12}
 & \multirow{2}{*}{2-2} & equilibrium-lcfs\_r & $(170,1,5)$ & 20 & 42.50 & 0.999 & 0.073 & 0.064 & \textbf{0.049} & 0.052 & \underline{0.050} \\
 &  & equilibrium-lcfs\_z & $(170,1,5)$ & 27 & 31.48 & 1.000 & 0.060 & 0.049 & \textbf{0.036} & 0.038 & \underline{0.037} \\
\cmidrule(lr){2-12}
 & \multirow{1}{*}{2-3} & equilibrium-psi & $(65,65,5)$ & 191 & 110.60 & 0.999 & 0.136 & 0.126 & \underline{0.094} & 0.097 & \textbf{0.092} \\
\midrule
\multirow{10}{*}{\rotatebox[origin=c]{90}{\textbf{Group~3}}} & \multirow{2}{*}{3-1} & thomson\_scattering-n\_e & $(120,1,10)$ & 456 & 2.63 & 0.999 & 0.339 & 0.314 & \textbf{0.263} & 0.288 & \underline{0.278} \\
 &  & thomson\_scattering-t\_e & $(120,1,10)$ & 927 & 1.29 & 0.999 & 0.453 & 0.434 & \textbf{0.408} & 0.422 & \underline{0.414} \\
\cmidrule(lr){2-12}
 & \multirow{3}{*}{3-2} & soft\_x\_rays-horizontal\_cam\_lower & $(18,1,2500)$ & 4474 & 10.06 & 0.999 & \textbf{0.242} & 0.291 & 0.255 & \underline{0.254} & 0.255 \\
 &  & soft\_x\_rays-horizontal\_cam\_upper & $(18,1,2500)$ & 8192 & 5.49 & 0.996 & 0.230 & 0.232 & \textbf{0.177} & 0.187 & \underline{0.179} \\
 &  & spectrometer\_visible-filter\_spectrometer\_dalpha\_voltage & $(3,1,2500)$ & 1763 & 4.25 & 0.999 & 0.520 & 0.483 & \textbf{0.364} & 0.396 & \underline{0.372} \\
\cmidrule(lr){2-12}
 & \multirow{5}{*}{3-3} & equilibrium-beta\_normal & $(1,1,1)$ & 1 & 1.00 & 1.000 & 0.336 & 0.289 & \textbf{0.188} & 0.202 & \underline{0.191} \\
 &  & equilibrium-beta\_pol & $(1,1,1)$ & 1 & 1.00 & 1.000 & 0.328 & 0.294 & \textbf{0.192} & 0.202 & \underline{0.193} \\
 &  & equilibrium-beta\_tor & $(1,1,1)$ & 1 & 1.00 & 1.000 & 0.303 & 0.297 & \textbf{0.173} & 0.188 & \underline{0.176} \\
 &  & thomson\_scattering-n\_e & $(120,1,1)$ & 57 & 2.11 & 0.999 & 0.557 & 0.374 & \textbf{0.268} & 0.289 & \underline{0.269} \\
 &  & thomson\_scattering-t\_e & $(120,1,1)$ & 106 & 1.13 & 0.999 & 0.406 & 0.340 & \textbf{0.263} & 0.272 & \underline{0.269} \\
\midrule
\multirow{8}{*}{\rotatebox[origin=c]{90}{\textbf{Group~4}}} & \multirow{2}{*}{4-1} & soft\_x\_rays-horizontal\_cam\_lower & $(18,1,5000)$ & 8192 & 10.99 & 0.999 & 0.266 & \underline{0.214} & 0.255 & 0.214 & \textbf{0.204} \\
 &  & soft\_x\_rays-horizontal\_cam\_upper & $(18,1,5000)$ & 8192 & 10.99 & 0.994 & 0.284 & 0.241 & 0.277 & \underline{0.233} & \textbf{0.223} \\
\cmidrule(lr){2-12}
 & \multirow{2}{*}{4-2} & soft\_x\_rays-horizontal\_cam\_lower & $(18,1,5000)$ & 8192 & 10.99 & 0.999 & 0.261 & \underline{0.215} & 0.264 & 0.222 & \textbf{0.205} \\
 &  & soft\_x\_rays-horizontal\_cam\_upper & $(18,1,5000)$ & 8192 & 10.99 & 0.994 & 0.286 & 0.242 & 0.287 & \underline{0.239} & \textbf{0.223} \\
\cmidrule(lr){2-12}
 & \multirow{1}{*}{4-3} & equilibrium-magnetic\_axis\_z & $(1,1,20)$ & 13 & 1.54 & 0.999 & 0.129 & 0.118 & 0.141 & \underline{0.095} & \textbf{0.080} \\
\cmidrule(lr){2-12}
 & \multirow{1}{*}{4-4} & summary-ip & $(1,1,400)$ & 29 & 13.79 & 0.999 & 0.530 & 0.404 & 0.420 & \underline{0.384} & \textbf{0.349} \\
\cmidrule(lr){2-12}
 & \multirow{2}{*}{4-5} & magnetics-b\_field\_pol\_probe\_omv\_voltage & $(3,256,98)$ & 8192 & 9.19 & 0.745 & 0.638 & \textbf{0.632} & 0.649 & 0.635 & \underline{0.634} \\
 &  & magnetics-b\_field\_tor\_probe\_cc\_field & $(3,256,98)$ & 8192 & 9.19 & 0.757 & \textbf{0.663} & \underline{0.665} & 0.673 & 0.666 & 0.668 \\
\bottomrule
\end{tabular}
}
\label{tab:per_signal_all_groups}
\end{table*}

%
%

\end{document}